\documentclass[aps,twocolumn,prl,showpacs]{revtex4}

\usepackage{amsbsy}
\usepackage{graphicx}
\begin{document}
\newcommand{\beq}{\begin{equation}}
\newcommand{\eeq}{\end{equation}}
\newcommand{\bear}{\begin{eqnarray}}
\newcommand{\eear}{\end{eqnarray}}
\newcommand{\rd}{\mathrm{d}}
\newcommand{\lp}{\left}
\newcommand{\rp}{\right}
\newcommand{\bs}{\mathbf}
\newcommand{\parl}{\parallel}
\newcommand{\bz}{\bs{\hat{z}}}
\newcommand{\bx}{\bs{\hat{x}}}
\newcommand{\by}{\bs{\hat{y}}}
\newcommand{\bn}{\bs{\hat{n}}}
\newcommand{\hb}{}
\newcommand{\G}{\Gamma}
\newcommand{\ex}{\eta_x}
\newcommand{\ey}{\eta_y}
\newcommand{\eo}{\eta_0}
\newcommand{\pd}{\partial}
\renewcommand{\b}{\beta}
\renewcommand{\a}{\alpha}
\renewcommand{\d}{\delta}
\newcommand{\D}{\Delta}
\newcommand{\tr}{\textrm}
\newcommand{\mr}{\mathrm}
\newcommand{\tm}{\times}
\newcommand{\tetr}{D_{4h}}
\newcommand{\be}{\vec{\eta}}
\newcommand{\bdp}{\bs{D}_{x,y} }
\newcommand{\bl}{\bs{l}}
\newcommand{\lpp}{l_\perp}
\newcommand{\blpp}{\bs{l}_\perp}
\newcommand{\bv}{\bs{v}}
\newcommand{\bk}{\bs{k}}
\newcommand{\bd}{\bs{d}}
\renewcommand{\bm}{\bs{m}}
\newcommand{\etl}{{\it et al.}}
\newcommand{\ruth}{$\tr{Sr}_2\tr{RuO}_4 $}
\setlength\arraycolsep{0pt}

\title{Orbital magnetic dynamics in chiral {\it p}-wave superconductors}

\author{V.~Braude and E.~B.~Sonin}
\affiliation{Racah Institute of Physics, The Hebrew University of
Jerusalem,
Jerusalem 91904, Israel}
\date{\today}
\begin{abstract}
We present a theory of orbital magnetic dynamics for a chiral
{\it p}-wave superconductor
 with broken time-reversal symmetry. In contrast to the common Landau-Lifshitz theory
 for spin ferromagnets,
 %in the case of orbital magnetism,
 %it is impossible to define unambiguously a spontaneous magnetic moment:
 %the latter depends on conditions of its experimental investigation.
 the case of orbital magnetism cannot be described in terms of  local magnetization density.
Hence it is impossible to define unambiguously the spontaneous
magnetic moment: the latter would depend on conditions of its
experimental investigation.
 As an example of this we consider
 orbital magnetization waves and the  domain structure energy.
 %Combining the unusual magnetic
 %orbital dynamics with the London electrodynamics for the superconducting order parameter, we derived
 %the spectrum of orbital magnetization waves.
\end{abstract}
\maketitle
 Superconductors with unconventional pairing mechanisms have attracted a lot of attention during the past
 decade. A prominent feature characteristic of unconventional superconductivity is the possibility of
 states with
 broken time-reversal symmetry (TRS), which is expressed in the presence of  magnetic structure in these
 materials. Broken TRS has been detected in several superconductors, including $\tr{Sr}_2\tr{RuO}_4$
 ,\cite{luke}  $\tr{ZrZn}_2$
 ,\cite{Pfleiderer}
 and  $\tr{UGe}_2$, \cite{Saxena} which brought renewed interest to the old-standing problem of
 coexistence of
 superconductivity
 and magnetism.

The layered perovskite $\tr{Sr}_2\tr{RuO}_4$ is one of the best
known examples of unconventional superconductors with broken TRS.
\cite{mac} Experimental observations of a temperature-independent
Knight shift  for $H \perp \bz$ (Ref.~\onlinecite{Ishida}) and an
increased muon spin relaxation below $T_c$
(Ref.~\onlinecite{luke}) support the theoretically proposed
spin-triplet
{\it p}-wave
order parameter \cite{Rice, Agterberg, Sigrist}
$\Delta_{\a\b}(\bk)=i \,d^i(\bk) (\sigma^i \sigma^y)_{\a\b}$, with
$\bd(\bk)=\D_0 \bz(k_x+ik_y) $. Such an order parameter has a
non-zero  orbital moment
%$\bl=-i \langle d_i^*(\bk) |\bk \tm \nabla_\bk| d_i(\bk) \rangle /\langle d_j^* | d_j \rangle$,
\beq \bl=-i \langle d_i^*(\bk) |\bk \tm \nabla_\bk| d_i(\bk)
\rangle /\langle d_j^* | d_j \rangle, \eeq which should lead, in
principle, to spontaneous magnetization like in usual
ferromagnets. However,
 coexisting
superconductivity screens out this magnetization, making its
experimental detection rather difficult.  As a possible way to
overcome this difficulty, it was proposed \cite{BS} to perform
microwave response measurements, where excitation of spin waves
would provide an immediate signature for the presence of magnetic
order.

Previous works \cite{Ng, BS, br, radzih, belitz} which considered
magnetic dynamics in superconductors with coexisting magnetism
(SCFM), assumed a phenomenological model in  which the
magnetization was independent from  the superconducting order
parameter, and its dynamics was described by the Landau-Lifshitz
equation (LL dynamics). \cite{LL} This model, however, is not so
obvious for  materials such as $\tr{Sr}_2\tr{RuO}_4$, where TRS is
broken at the superconducting transition, and the magnetic
properties are expected to stem from the orbital part of the
multicomponent superconducting order parameter.
%The following features of these systems contradict the basic assumptions
%of the LL SCFM's: 1. Magnetic properties are not due to spin but due to orbital moment of Cooper pairs;
%2. In contrast to the LL SCFM's, where the spin moment density is a well-defined quantity, the orbital
%magnetization density
Therefore a proper treatment of magnetic dynamics in such
materials should derive it from the dynamics of the
superconducting order parameter. An analogous derivation has been
done \cite{wolfle} for the A phase of superfluid $^3\mr{He}$
($^3$He-A); then a modification of the theory taking into account
the charge of Cooper pairs would yield magnetic orbital dynamics
for an isotropic
 {\it p}-wave superconducting electron liquid.
However, such a derivation cannot be applied directly to the case
of a superconducting metal because of the crystal-field
anisotropy.

In this paper, we address this problem and derive an effective
orbital magnetic dynamics for a {\it p}-wave superconductor in a
strong crystal-field potential. The resulting dynamics is not
equivalent to the phenomenological LL SCFM magnetization dynamics,
except for some simple cases. Moreover, in general, it cannot be
described in terms of local magnetization at all! Instead, we show
that it can be described with the help of a unit vector $\bl$,
corresponding to the angular momentum of Cooper pairs. A similar
situation happens in $^3\tr{He}$ A phase, where the presence of an
anomalous term in the current does not allow us to express the
dynamical equations in terms of local angular momentum density.
\cite{vol,orb}

 We start by considering a Ginzburg-Landau (GL) free energy functional for a {\it p}-wave superconductor in
 a strong crystal field possessing a tetragonal symmetry $\tetr$ (which is the symmetry  of
 $\mr{Sr}_2\mr{RuO}_4$).
 %Spin dynamics for such materials has been considered in the past \cite{kee}.
 For simplicity, we will assume that the spin does not participate in the dynamics.
  This assumption is justified, for example,
 in the case when  the direction of $\bd(\bk)$ is fixed  along the crystal axis $\bz$ by   strong
 spin-orbit coupling. The order parameter can be decomposed into five odd-parity irreducible
representations of $D_{4h}$ symmetry. However, limiting the order
parameter to a {\it p}-wave form and requiring $\bd \parl \bz$
leaves only two of them:  a two-dimensional $\G_5^-=\{\bz k_x, \bz
k_y\}$,
 and a one-dimensional $\G_1^-= \{\bz k_z\}$. Having in mind  the order parameter
 $\sim k_x+ik_y$ for
$\mr{Sr}_2\mr{RuO}_4$,
we assume
that the $\G_5^-$ representation is most favorable, having
the highest transition temperature. It will be described by the GL order parameter
$\be=(\eta_x, \eta_y)$. However, this representation alone
is not enough to describe the magnetic properties, since it does not allow any magnetic moment
 perpendicular to $\bz$, and hence we will need an admixture of the
higher-energy representation $\G_1^-$. The free energy then has the form
\beq \label{eq:general}
  f=f_5+f_1+f_{1-5}+b^2/8\pi,
\eeq where $\bs{b}=\nabla\tm \bs{A}$ is the magnetic field. The
form of the GL functional corresponding to representations
$\G_5^-$ and $\G_1^-$ is well-known: \cite{ueda} \bear
\label{eq:f5hom}
 f_{5 \,hom }&=&P_1(T)|\be|^2+\b_1|\be|^4+\b_2(\eta_x^* \eta_y-\eta_x \eta_y^*)^2
    \nonumber \\&&+\b_3|\eta_x|^2 |\eta_y|^2,
\eear
 for the homogeneous part of the functional corresponding to the $\G_5^-$ representation,
\bear  \label{eq:f5grad}
 &&f_{5 \,grad }=K_1|\bdp \cdot \be|^2+K_2[|D_x \eta_y|^2+|D_y \eta_x|^2]
     \nonumber \\ &&+K_3[(D_x \eta_x)^* (D_y \eta_y)+\tr{c.c.}]+K_4[(D_x \eta_y)^* D_y \eta_x+
     \tr{c.c.}]
   \nonumber \\ &&+K_5 [|D_z \eta_x|^2+|D_z \eta_y|^2]
  \eear
 [where $D_i\equiv \pd_i-i(e^*/c\hbar) A_i$] for the gradient part of the $\G_5^-$ functional, and
  \bear
    f_1&=&P_2(T) |\psi|^2+K_6(|D_x \psi|^2+|D_y \psi|^2)
    \nonumber \\ &&+K_7 |D_z \psi|^2
    \eear
    for the functional corresponding to the $\G_1^-$ representation.
For $P_1(T)<0$ and $P_2(T)>0$, the ground state is determined by $f_{hom \,5}$, which gives
  $\be=\eta(1,\pm i)$,
provided that $\b_2>0$, $4\b_2>\b_3$, and $4(\b_1-\b_2)+\b_3>0$. This reproduces the order parameter for
\ruth, and   it is also
%TRS-breaking state is
analogous to the Anderson-Brinkman-Morel (ABM) state in superfluid
$^3\mr{He}$
 with the $\bs{l}$ vector parallel to the $\bz$ axis
\cite{wolfle}. Assuming the couplings of $f_{5 \, hom}$ to be
dominant (consistent with the strong anisotropy and the tendency
to TRS breaking), the order parameter $\be$ is frozen in the
ground state \beq \label{eq:gs} \be=\eta(1,i) \quad\tr{and} \quad
|\eta|^2 \equiv \eta_0^2=\frac{-P_1(T)}{4(\b_1-\b_2)+\b_3}, \eeq
while the low-lying excitations are described by soft variables:
$\psi$ and the  total  phase
  $\varphi$.

The mixing terms between $\G_5^-$ and $\G_1^-$ are determined by
requirements of invariance under the point group, as well as under
the gauge transformation and the time reversal.
%We write down such terms which do not vanish for $\eta$ given by Eq.~(\ref{eq:gs}).
The lowest-order homogeneous mixing terms are
\beq
 |\be|^2 |\psi|^2 \oplus \be\,^2 (\psi^*)^2+\tr{c.c.}
\eeq
 The gradient mixing terms are taken to second order in $\psi$. Then there are 2 types of such
 terms: terms such as $\sim (D \eta)^* D \psi$ and such as $\sim (\eta D \psi)^2$. Writing only those terms which
 do not vanish for the state given by Eq.~(\ref{eq:gs}), we have
 two terms of   the first type:
\beq
   \lp[(D_z \psi)^*  \bdp \cdot \be   +\tr{c.c.}
   \rp]\oplus \lp[ D_z^* \be^* \cdot \bdp \psi+\tr{c.c.} \rp],
 \eeq
 while for the second type four terms are possible:
\bear   &&|\be|^2|\bdp \psi|^2 \oplus \eta_x^* \eta_y^* D_x \psi
D_y \psi+\tr{c.c.}   \oplus
   \nonumber \\ &&
  (\eta_x \eta_y^*-\eta_y \eta_x^*)[(D_x \psi)^* D_y \psi-
  (D_y \psi)^* D_x \psi] \oplus
  \nonumber \\ &&(\eta_x^2-\eta_y^2)^*\lp[(D_x \psi)^2-(D_y
  \psi)^2\rp]+\tr{c.c.}
  \eear
 Collecting all terms and using Eq.~(\ref{eq:gs}), we obtain for the free energy:
\bear  \label{eq:energy}
 && f=(K_1+K_2)|\bdp \eta|^2+(K_3-K_4)i[(D_x \eta)^* D_y \eta-\tr{c.c.}]   \quad
   \nonumber \\&&+2K_5|D_z \eta|^2+
  Q_1[(D_z \psi)^* (D_x+iD_y)\eta+\tr{c.c.}]+   \nonumber \\
 && Q_2[(D_z \eta)^* (D_x-iD_y)\psi+\tr{c.c.}]  +
  Q_3 i[(D_x \psi)^* D_y \psi-\tr{c.c.}]
  \nonumber \\&&+Q_4 \lp\{(\eta^*)^2[
  (D_x \psi)^2-(D_y \psi)^2]+\tr{c.c.} \rp\}+
  \nonumber \\ &&
  Q_5 i[ (\eta^*)^2 D_x \psi D_y \psi  -\tr{c.c.}]
   +f_1+P_3 \eta_0^2 |\psi|^2+b^2/8 \pi.
\eear
    As we have already mentioned, the state
  $\psi=0, \be=\eta(1,i)$ corresponds to the ABM state with $\bl \parl \bz$.
  We would like to exploit this analogy and express the order parameter in terms of the unit
  vector $\bl$.
  Since the triple
  $(\eta_x, \eta_y, \psi)$ transforms like a vector under rotations, we can produce a state with
  nonzero $\psi$ by making a small rotation around an axis in the x-y plane. Indeed,
  \beq
    R_{\d \a, \,\bs{n}} \eta (1, i ,0)=\eta(1,i,-\d\a e^{i \phi}),
  \eeq
  where   $\d\a$ is a small rotation angle, and $\phi$ the angle between the axis of rotation $\bn$
  and $\by$  axis. From this we identify $\psi/\eta=-\d\a\, \exp[i\phi]$ . On the other hand,
  $l_z \d\a  \,\exp[i\phi] =(l_x+i l_y) $, so we can make a replacement $\psi=-\eta(l_x+il_y)$. Thus
  the presence of a small admixture of the order parameter $\psi$ is analogous to tilting the
  vector $\bl$
  away from the $\bz$ direction. We stress that this is only correct as long as $\psi$ is
  small. Large deflections of $\bl$ require a breaking of  the state given by Eq.~(\ref{eq:gs}),
  and then the configuration space $(\be, \psi)$ might include also states in which $\bl$ is not
  defined
  at all (TRS-conserving states). This is a consequence of strong crystal-field anisotropy
  (for  a weak anisotropy, like in $^3\tr{He}$, this would not be a problem, since there states with
  defined $\bl$ - namely, the ABM states -  are well separated by the energy from other states).

In addition to the tilting of the vector $\bl$,  excitation of the
orbital mode results also in superconducting flow with the
velocity $  \bs{v} = \nabla \varphi/ m^*-( e^*/ m^* c) \bs{A}$.
% $\bs{D} \eta = i \eta m^* \bv$.
%variation of the total phase $\varphi$ of the order parameter. The phase gradient determines the
%gauge-invariant velocity: $  \bs{v} = (\hbar /m^*) \nabla \varphi -(\hbar e/ m^* c) \bs{A}$.
We would like to mention
  in passing that
  since only small deviations of $\bl$ from $\bz$ are considered, no problem of nonzero
  %$\nabla \tm \bs{v}$
  curl of the phase gradient (Mermin-Ho relation \cite{ho}) arises here. Expressing
  Eq.~(\ref{eq:energy}) in terms of the new variables
  $ \blpp=(l_x, l_y)$, and $\bv$ (in the harmonic approximation), we obtain for the free energy
\bear \label{eq:l-energy}
  && f=\frac{1}{2}\rho_{xy} |v_{xy}|^2+\frac{1}{2}\rho_z |v_z|^2+B_1 |\nabla \cdot \blpp|^2
\nonumber \\ &&
  +B_2 |\bz  \cdot \nabla \tm \blpp|^2
  %\nonumber \\ &&
   +B_3 |\pd_z \blpp|^2+
  C\, \bv \cdot \nabla \tm \blpp
\nonumber \\ &&
  -C_{an} \bv_z \cdot
  \nabla \tm \blpp    %B_1 |\nabla \cdot \blpp|^2+B_2 |\bz  \cdot \nabla \tm \blpp|^2+
 % \nonumber \\ &&
 % B_3 |\pd_z \blpp|^2+
 %\nonumber \\ &&
  +Z \pd_x l_x \pd_y l_y +\frac{\a}{2} |\lpp|^2+b^2/8\pi, \quad
\eear where all the coefficients may be expressed via those of
Eq.~(\ref{eq:energy}): for example, $\a=2[P_2(T) \eta_0^2+P_3
\eta^4]$, $C=Q_1 \eta_0^2 m^*$, and $C_{an}=(Q_1+Q_2)\eta_0^2
m^*$. Without anisotropy terms $\propto \alpha$ and $\propto Z$,
this functional is analogous to the energy of the $^3$He-A phase,
\cite{wolfle} linearized with respect to small deviations from the
$\bz$ axis.  The free energy determines the (charge) current
density: \beq \label{eq:current0}
  \bs{j}=  \frac{e^*}{m^*} \frac{\pd f}{\pd v}=  \bs{j}_{tr} +  \bs{j}_m  ~,
     \eeq
where
%$\bs{j}_{tr}= (e^*/m^*)( \rho_{xy} \bv_{xy}+  \rho_z  \bv_z )$
\beq \label{eq:current}
   \bs{j}_{tr}= \frac{e^*}{m^*}( \rho_{xy} \bv_{xy}+  \rho_z  \bv_z )
\eeq
is the transport current, and
\beq \label{eq:current1}
  \bs{j}_m= \frac{e^*}{m^*}( C \nabla \tm \blpp-C_{an} \nabla_{xy} \tm \blpp)
\eeq
is the magnetization  current [here $\nabla_{xy}\equiv (\pd_x, \pd_y,0)$].

Comparing  with the LL SCFM model, \cite{Ng,BS} one can see that
it differs from the latter not only by anisotropy in the London
penetration depth $\sim 1/\rho_i$ and the in-plane term  $\sim Z$,
but, most importantly, by the presence of an anomalous term $\sim
C_{an}$ in the magnetization current. This term  renders it
impossible to have a consistent definition of magnetization for
the system (despite the presence of magnetization currents!),
which makes a crucial difference between our {\it p}-wave
superconductor and
 LL SCFM's.
Indeed, this current cannot be expressed as a curl of magnetization vector $c \nabla \tm \bm$,
%$ \bs{j}_m=c \nabla \tm \bm$.
 and therefore  the energy of interaction
$(m^*/e^*)\bs{v}\cdot   \bs{j}_m$   cannot be cast in the Zeeman form $-\bs{b}\cdot\bm$.
%As a result, it is impossible to have a consistent definition of magnetization for the system,
%identify the vector $\bl$ as a magnetization,
%which makes a crucial difference between our {\it p}-wave superconductor and
%LL SCFM's.
 We emphasize here that the anomalous term appears not due to the
anisotropy, but, rather, due to the orbital nature of magnetism in
our system.

The magnetic dynamics is generally described by  Ampere's law
\beq
  \frac{4 \pi}{c} \lp(\bs{j}_{tr}+\bs{j}_m \rp)=\nabla \tm b~,
\eeq by the London equation \beq \label{eq:london}
  \nabla \tm \bv=-\xi \bs{b}~,
\eeq
 where $\xi \equiv (e^*/m^* c )$, and by an equation of motion for $\bl$.
This motion is described by the equation
 \beq \label{eq:prec}
  \frac{\pd \blpp}{\pd t}=-g \bz \tm \frac {\d f}{\d \blpp}=-g \bz \tm\lp(\frac{\pd f}{\pd \blpp}-\pd_i
  \frac{\pd f}{\pd \pd_i \blpp} \rp),
\eeq which is a precession equation for a unit axial vector, with
a generalized torque given by the expression in the brackets. This
equation can be obtained from thermodynamic conservation
requirements by a general procedure used for the derivation of
hydrodynamic equations. \cite{LL6} Its generalization for
%It is a simplified version of
the orbital dynamics
%equation for
of $^3\tr{He-}A$  has been considered in Ref.~\onlinecite{hu}
[Eq.~(\ref{eq:prec}) is obtained by assuming small deviations from
the equilibrium and zero normal-fluid velocity].
 Although
this equation formally looks the same as the LL equation for spin
ferromagnets, there is an essential difference: the dynamical
constant $g$ here is an
 unknown phenomenological parameter, while in the LL equation
it is given by $g=\gamma/M_0$, where $\gamma$ is the gyromagnetic
ratio, and $M_0$ is the magnetization. The value of $g$ was
disputed for $^3$He-A, \cite{vol,orb} but its determination is a
prerogative of the microscopic theory.

Explicitly the
%precession
 equation of motion for the $\bl$ vector  reads
\bear \label{eq:l-dyn}
 &&\dot{\bl}_\perp=-g \bz \tm \big[\a \blpp-2B_1 \nabla (\nabla \cdot \blpp)
 -2B_2 \bz \tm \nabla (\nabla \tm \blpp)_z
 \nonumber \\ &&
 -2 B_3
  \pd_z^2 \blpp
%  \nonumber \\ &&
  -C \xi \bs{b}- C_{an} \nabla \tm \bv_z-Z \vec{\pd}_j \pd_i l_i (1-\d_{ij}) \big]. \quad
\eear
 Equations~(\ref{eq:current0})-(\ref{eq:london}) and (\ref{eq:l-dyn}) constitute a full system of
equations describing the magnetic dynamics. As was explained
above, the magnetic order parameter $\bl$ cannot be identified as
a local magnetization, despite being analogous to it, and hence
the dynamics is, in general, more complicated than the LL SCFM
dynamics. In special cases, however, namely in situations where
all variations are either parallel or perpendicular to the $\bz$
axis, the dynamics can be described in terms of local
magnetization, so an effective LL SCFM description is valid (with
a generalization of using tensorial stiffness parameters and the
London penetration depth, as well as in-plane anisotropy). Still,
the value of the effective magnetization determined this way would
not be unique, but rather different in each case.
 For example, for
magnetization waves, propagating in the $\bz$ direction
(perpendicular geometry), the free energy is \beq
  f= \frac{\rho_{xy}}{2} |v_{xy}|^2 -C \xi \bs{b}\cdot \blpp +B_3 |\pd_z \blpp|^2 +\frac{\a}{2}
  |\lpp|^2+\frac{b^2}{8\pi} ,
\eeq leading for plane waves $\propto e^{i \bs{q} \cdot
\bs{r}-i\omega t}$ to a dispersion \beq
  \frac{\omega}{g}=\pm \lp(\a+2B_3 q^2-4\pi \frac {C^2 \xi^2 q^2}{q^2+4\pi \rho_{xy}\xi^2} \rp).
\eeq This has  a form of a LL SCFM spectrum that was considered in
Ref. \onlinecite{BS} with an equilibrium magnetization $M_0 =\xi
C$.
%with the identification of parameters:
%\bear
%&&  \rho_{xy}\to \frac{1}{4\pi} \lp(\frac{m^* c}{e^* \lambda} \rp)^2; \quad C\to
%\frac{ m^* c M_0}{e^*}; \quad
%\nonumber \\ &&
%\a\to 4 \pi \a M_0^2; \quad
 %\nonumber \\ &&
% B_3 \to 2\pi \a l_d^2 M_0^2 ,
%\eear
 Hence the results for the microwave response of a LL SCFM given in that paper
are directly applicable to the case of an unconventional
superconductor considered here,
%(assuming that a simple
%boundary condition  $\pd_z \bl=0$ is valid here).
%Note that in this case the dynamics  can be described in terms of an equilibrium
provided that an effective equilibrium magnetization is chosen as
$M_0=\xi C$.
%proportional to the coefficient $C$, and the
%London penetration depth is determined by the coefficient $\rho_{xy}$.
 On the other hand, for waves propagating in the $\bx-\by$ plane, the spectrum is
\bear
 && \frac{\omega^2}{g^2}=-\lp(\frac{Z q^2 \sin4\phi}{4}\rp)^2+\lp[\a+(2B_1+\frac{Z \sin^2
 2\phi}{2})q^2\rp]
\nonumber \\ && \, \tm \lp[\a+(2B_2-\frac{Z \sin^2 2\phi}{2})q^2
-4\pi \frac{(C-C_{an})^2\xi^2 q^2}{q^2+ 4\pi \rho_z \xi^2}\rp],
\eear
where $\phi$ is the angle the wave vector $\bs{q}$ makes
with the $\bx$ axis. This expression corresponds to the LL SCFM
spectrum  in the parallel geometry \cite{br} with a different
value of equilibrium magnetization, $M_0=\xi (C-C_{an})$
  (in that analysis no $\bx-\by$ plane anisotropy was introduced, so
the $\phi$ dependence was irrelevant there). We stress again,
however, that these results  cannot be interpreted in terms of a
tensorial $M_0$. For a general propagation direction the dynamics
cannot be interpreted in terms of local magnetization at all.
%This expression is complicated (in comparison
%to the SCFM)
 %by an in-plane
%anisotropy, and by the unequal in-plane longitudinal and transverse stiffness.

\begin{figure}
\begin{center}
\includegraphics[width=0.4\textwidth]{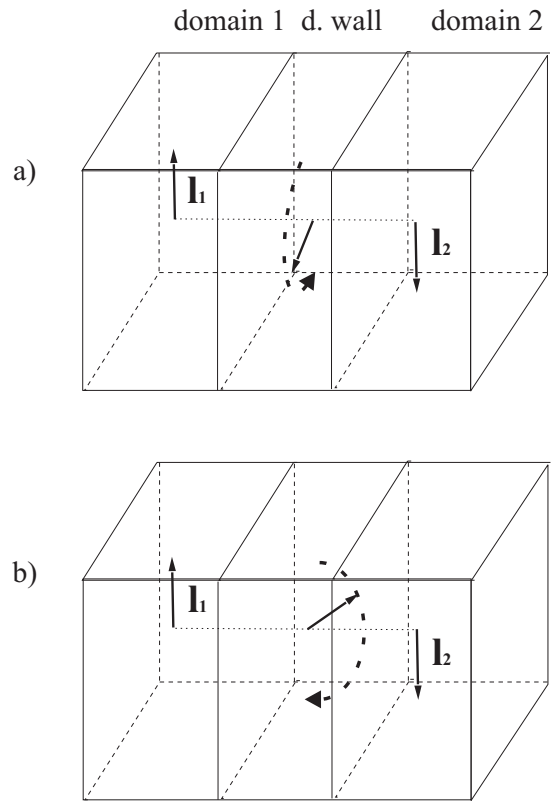}
\end{center}
\caption{Domain wall between two domains with opposite directions
of the $\bl$ vector in the case of weak anisotropy: a)Bloch wall;
b)N\'eel wall. Note that when the anisotropy is strong $\bl$ is
not defined inside the wall.} \label{domain}
\end{figure}

Another issue in which important differences arise between our
unconventional superconductor and a LL SCFM, is the  issue of
domain walls. A basic feature of ferromagnets, both insulating and
superconducting, which determines the field distribution inside
domains, is the  discontinuity of  the magnetic induction $ \D
b_z$ across the wall being given by $8 \pi M_0$, with $M_0$ the
equilibrium magnetization inside the domains. Moreover, this
discontinuity may be used as an experimental definition of domain
magnetization. One might ask, whether in our case $\D b_z/8 \pi$
has any meaning of magnetization and whether it is related in any
sense to dynamic response properties discussed above. The answer
to the second question is, in general, negative.
%Namely, $\D B_z$ is not related to  $C$, $C_{an}$, or $g$.
Indeed, in order to determine $\D b_z$, one has to examine what
happens inside the domain wall, which involves a strong deviation
from the ground state, Eq.~(\ref{eq:gs}). For that purpose, the
free energy  Eq.~(\ref{eq:l-energy}) is inapplicable, and one has
to go back to the general expression, Eq.~(\ref{eq:general}). With
the  energy $f_{5\,hom}$ being dominant, a domain-wall solution
may be restricted to the $(\eta_x, \eta_y)$
 part of the order parameter.
For  specific parameter values, domain-wall solutions have been
obtained, \cite{volovik, sigrist2} with the discontinuity being
determined by the parameters of $f_{5 \, grad}$. For example, when
$\b_1, \b_3 \gg \b_2$, and $K_3=K_4$, the discontinuity is
\cite{volovik} \beq \D b_z=\frac{4 \pi}{c} \frac{P_1(T) e^*
K_3}{\b_3} \frac{K_1-K_2}{K_1+K_2}. \eeq Thus, $\D b_z$ has
nothing to do either with the parameters $C$ and $C_{an}$, which
determine the magnetization current for small deviations, or the
dynamical parameter $g$. Only in the limit of very  weak crystal
anisotropy, when the vector $\bl$ is well defined even inside the
domain wall  as shown in Fig.~\ref{domain}, the discontinuity $\D
b_z$ is determined by a combination of $C$ and $C_{an}$. Indeed,
in this case
 the
magnetization current inside a domain wall is given by the
isotropic form
\beq
     \bs{j}_m= \frac{e^*}{m^*}\lp[ C \nabla \tm \bl-C_{an} \bl (\bl \cdot \nabla \tm \bl) \rp],
\eeq
%where it rotates far from the axis $\bz$,
%the discontinuity $\D b_z$ is determined by a combination of $C$ and $C_{an}$.
%Thus,
and then the field discontinuity for a Bloch wall (where $\bl$
rotates in the wall plane) is given by $\D b_z=8 \pi \xi
(C-C_{an})$, corresponding to an effective magnetization
$\xi(C-C_{an})$. On the other hand, for a N\'eel wall (where $\bl$
rotates in the plane perpendicular to the wall) the discontinuity
becomes $\D b_z=8\pi \xi C$, corresponding to  an effective
magnetization $\xi C$.
%The discontinuity
%becomes $8\pi \xi C$ for the Neel wall, in which $\l$ rotates in the plane normal to the wall plane.

As to the first question,  $\D b_z/8 \pi$  does have the
correspondence to magnetization in the usual ferromagnets. To see
it, let us recall that for usual ferromagnets
 the domain-structure energy is given by the sum of a local domain-wall
energy (which may be considered an effective parameter independent of the fields)
 and an electromagnetic energy inside the domains (including the energy of Meissner currents
and the Zeeman term $-b_z M_0$).
In our case, the Zeeman term is absent, and instead the
domain-wall energy contains a  contribution due to the domain-wall currents.
However, for thin domain walls this contribution can be transformed to an effective Zeeman form:
%Though the Zeeman term is absent in general, it can reappear in the limit of very thin domain wall
%(compared with the London penetration depth and the domain size). In this limit one can neglect
%variation of the electromagnetic potential $\bs{A}$ inside the domain wall, and the energy of
%interaction between the magnetization and transport currents can be transformed  to the Zeeman-type
%form:
\beq
  -\int_{wall} \frac{j_y A_y}{c} \,d x=- \int_{domain} \tilde M b_z(x) \, d
  x,
\eeq where $\tilde M= \D b_z/8\pi$, and the domains are in the
$\bx$ direction. Hence $\D b_z/8\pi$ may be interpreted as an
effective magnetization of each domain, which allows the
electromagnetic part of the domain-wall energy to be cast in the
usual Zeeman form, so that the remaining part is entirely local
(i.e., independent of the field distribution). This static
``magnetization'' determines the field in the domains, but is
unrelated, in general, to the dynamical response of the material.
As an application of this result, one can immediately obtain a
criterion of stability against the formation of domains  for our
material. For this purpose we can use the result of Krey's
analysis \cite{krey} for a LL SCFM, that the uniformly magnetized
configuration becomes unstable when $W<4\pi \lambda
 M_0^2$,
where $W$ is the local domain-wall energy (calculated excluding
electromagnetic effects), and $\lambda$ the London penetration
depth. The same result is valid for our case if $\tilde M$ defined
above is substituted for $M_0$, and  the in-plane penetration
depth is used, $\lambda^{-2} \to 4 \pi \xi^2 \rho_{xy}$. Of
course, $W$ should be calculated separately, as was done in
Refs.~\onlinecite{volovik} and  \onlinecite{sigrist2}.

In summary, we have studied the orbital magnetic dynamics in a
{\it p}-wave superconductor with strong crystal-field anisotropy.
The dynamics is essentially different from the Landau-Lifshitz
dynamics for a superconducting ferromagnet. The most important
difference is  that the directional order parameter $\bl$ (orbital
moment of Cooper pairs) does not lead to a definite spontaneous
magnetization (magnetic moment per unit volume).  While in simple
cases one can introduce an effective magnetic-moment density
similar to that in the Landau-Lifshitz dynamics, the value of this
density varies from case to case. As examples of these cases we
have considered  magnetization waves along and normal to the main
crystal axis, and the energy of the domain structure.
%criterion of stability against domain
%formation.

This work was supported by the Israel Academy of Sciences and
Humanities.


\begin{thebibliography}{99}
 %\bibitem{mineev} V.~P.~Mineev and T.~Champel, \prb {\bf 69}, 144521 (2004).
 %\bibitem{luke} G.~M.~Luke \etl, Nature (London), {\bf 394}, 558 (1998).
 \bibitem{luke} G.~M.~Luke, Y.~Fudamoto, K.~M.~Kojima, M.~I.~Larkin, J.~Merrin,
B.~Nachumi, Y.~J.~Uemura, Y.~Maeno, Z.~Q.~Mao, Y.~Mori,
H.~Nakamura, and M.~Sigrist, Nature (London), {\bf 394}, 558
(1998).
 %\bibitem{Pfleiderer} C.~Pfleiderer \etl, Nature (London) {\bf 412}, 58 (2001).
 \bibitem{Pfleiderer} C.~Pfleiderer, M.~Uhlarz, S.~M.~Hayden, R.~Vollmer,
 H.~von Lohneysen, N.~R.~Bernhoeft, and G.~G.~Lonzarich,
 Nature (London) {\bf 412}, 58 (2001).
 %\bibitem{Saxena} S.~S.~Saxena \etl, Nature (London) {\bf 406}, 587 (2000).
 \bibitem{Saxena} S.~S.~Saxena, P.~Agarwal, K.~Ahilan, F.~M.~Grosche,
 R.~K.~W.~Haselwimer, M.~J.~Steiner, E.~Pugh, I.~R.~Walker, S.~R.~Julian,
 P.~Monthoux, G.~G.~Lonzarich, A.~Huxley, I.~Sheikin, D.~Braithwaite,
 and J.~Flouqet,  Nature (London) {\bf 406}, 587 (2000).
\bibitem{mac} A.~P.~MacKenzie and Y.~Maeno, Rev.~Mod.~Phys. {\bf 75}, 657 (2003).
 \bibitem{Ishida} K.~Ishida, H.~Mukuda, Y.~Kitaoka, K.~Asayama, Z.~Q.~Mao, Y.~Mori, and Y.~Maeno,
  Nature (London), {\bf 396}, 658 (1998).
 \bibitem{Rice} T.~M.~Rice and M.~Sigrist, J.~Phys.~Condens.~Matter {\bf 7}, L643 (1995).
 \bibitem{Agterberg} D.~F.~Agterberg, T.~M.~Rice, and M.~Sigrist, \prl {\bf 78}, 3374 (1997).
 %\bibitem{Sigrist} M.~Sigrist \etl, Physica C (Amsterdam) {\bf 317}, 134 (1999).
\bibitem{Sigrist} M.~Sigrist D.~Agterberg, A.~Furusaki, C.~Honerkamp, K.~K.~Ng,
T.~M.~Rice and M.~E.~Zhitomirsky, Physica C (Amsterdam) {\bf 317},
134 (1999).
 \bibitem{Ng} T.~K.~Ng and C.~M.~Varma, \prb {\bf 58}, 11624(1998).
 \bibitem{BS} V.~Braude and E.~B.~Sonin, Phys.~Rev.~Lett.~{\bf 93}, 117001 (2004).
 \bibitem{br} V.~Braude, cond-mat/0601386, \prb (to be published).
 \bibitem{radzih} L.~Radzihovsky, A.~M.~Ettouhami, K.~Saunders, and J.~Toner, \prl {\bf 87},
 027001-1 (2001).
 \bibitem{belitz} S.~Tewari, D.~Belitz, T.~R.~Kirkpatrick, and J.~Toner, \prl {\bf 93}, 177002-1 (2004).
 \bibitem{LL} L.~D.~Landau, E.~M.~Lifshitz, and L.~P.~Pitaevskii, {\it Statistical Physics, Part 2}
 (Pergamon, Oxford, 1980).
 %\bibitem{kee} H.-Y. Kee, Y.~B.~Kim, and K.~Maki, \prb {\bf 61}, 3584 (2000).
\bibitem{wolfle} D.~Vollhardt and P.~W\"olfle, {\it The Superfluid phases of Helium 3} (Taylor \&
Francis, London, 1990).
\bibitem{vol} G. E. Volovik,  {\it Exotic Properties of Superfluid $^3$He} (World Scientific,
Singapore, 1992).
\bibitem{orb} E.B.
Sonin,  Physica B
{\bf 178}, 106 (1992).
\bibitem{ueda} M.~Sigrist and K.~Ueda, Rev.~Mod.~Phys.~ {\bf 63}, 239 (1991).
\bibitem{ho} N.~D.~Mermin and T.-L.~Ho, \prl {\bf 36}, 594 (1976).
\bibitem{LL6} L.~D.~Landau, E.~M.~Lifshitz, {\it Fluid mechanics}
(Pergamon, Oxford, 1987).
\bibitem{hu} C.-R. Hu and W.~M.~Saslow, \prl {\bf 38}, 605 (1977).
\bibitem{volovik} G.~E.~Volovik and L.~P.~Gor'kov, Zh.~Eksp.~Teor.~Fiz. {\bf 88}, 1412 (1985)
[Sov.~Phys. JETP {\bf 61}, 843 (1985)].
\bibitem{sigrist2} M.~Sigrist, T.~M.~Rice, and K.~Ueda, \prl {\bf 63},1727 (1989).
\bibitem{krey} U. Krey, Intern. J. Magnetism, {\bf 3}, 65 (1972).


 \end{thebibliography}
\end{document}